\documentclass[pra,aps,twocolumn,superscriptaddress,showpacs]{revtex4}

\usepackage{psfig}

\begin{document}

\title{Highly asymmetric quantum cloning in arbitrary dimension}

\author{Jarom\'{\i}r Fiur\'{a}\v{s}ek}
\affiliation{Department of Optics, Palack\'{y} University,
17. listopadu 50, 77200 Olomouc, Czech Republic}
\affiliation{QUIC, Ecole Polytechnique, CP 165,
Universit\'{e} Libre de Bruxelles, 1050 Bruxelles, Belgium }

\author{Radim Filip} 
\affiliation{Department of Optics, Palack\'{y} University, 
17. listopadu 50, 77200 Olomouc, Czech Republic}

\author{Nicolas J. Cerf\,}
\affiliation{QUIC, Ecole Polytechnique, CP 165,
Universit\'{e} Libre de Bruxelles, 1050 Bruxelles, Belgium }

\begin{abstract}
We investigate the universal asymmetric cloning of states
in a Hilbert space of arbitrary dimension. 
We derive the class of optimal and fully asymmetric
$1\rightarrow 3$ cloners, which produce three copies,
each having a different fidelity. A simple parametric expression 
for the maximum achievable cloning fidelity triplets is then provided.
As a side-product, we also prove the optimality of the $1\rightarrow 2$ 
asymmetric cloning machines that have been proposed in the literature. 
\end{abstract}

\maketitle

\section{Introduction}

Quantum information theory exploits the laws of quantum mechanics to devise
novel means of processing, manipulating and transmitting information. Among the
most celebrated applications one finds quantum computing and quantum
cryptography. The latter allows a secure key distribution among two distant
partners, the security of the distributed key being guaranteed by the laws of
quantum mechanics \cite{Gisin02}. In particular, the linearity of quantum mechanics 
implies that an unknown quantum state cannot be copied \cite{Wootters82}. 
Thus any attempt by an eavesdropper to learn about the state transmitted from 
the sender to the receiver will unavoidably introduce some noise, which can 
be detected at the receiver's station. 

Although perfect copying is forbidden it is still possible to carry out an
approximate cloning of quantum states. This issue has attracted a lot of
attention during the recent years and the optimal universal symmetric 
cloning machines which produce $M$ approximate copies out of $N$ inputs have 
been found \cite{Buzek96,Gisin97,Buzek98,Werner98,Bruss98,Buzek98flocks}. 
In the context of
quantum cryptography, one is particularly interested in the asymmetric cloning
machines which produce two clones with different fidelities
\cite{Cerf98,Niu98,Niu99,Cerf00,Braunstein01,Filip04}. This allows one to study
the interplay between the information gained by an eavesdropper and the noise
introduced in the channel. Importantly, the asymmetric cloning proved to be 
a very efficient (or even optimal) individual eavesdropping attack on certain
kinds of QKD protocols \cite{Fuchs97,Bechmann99,Cerf02,Bruss02}. Recently,
optimal asymmetric $1\rightarrow 2$ cloning of qubits encoded as polarization
states of single photons has been demonstrated experimentally \cite{Pan04}.

However, the universal asymmetric cloning  machines considered 
in the literature \cite{Cerf98,Niu98,Cerf00,Braunstein01} 
are only conjectured to be optimal, and so far the proof of optimality 
has been missing (except for the qubit case \cite{Niu98}). 
In this paper, we provide such a proof. We will then go beyond the
$1\rightarrow 2$ asymmetric cloning and shall consider a novel class of
universal asymmetric machines which produce three clones, each of them with
possibly different fidelity. These machines were recently proposed and briefly
discussed in Ref. \cite{Iblisdir04}  which introduced the general concept of 
a  fully asymmetric $N \rightarrow M$ cloning machine producing
$M$ approximate clones with $M$ different fidelities. In this paper we expand
this discussion and derive explicitly the optimal 
cloning transformation, present the details of the optimality proof, 
and provide a simple parametric description of the optimal
universal asymmetric $1\rightarrow 3$ cloning machines in arbitrary dimensions. 
We expect that our
findings will play an important role in investigations of multi-party quantum
communication protocols and quantum information distribution in quantum
networks. An independent similar study of multipartite asymmetric cloning of
qubits is reported in \cite{Iblisdir05}.

The paper is structured as follows. In Section II we prove the optimality of the
universal $1\rightarrow 2$ asymmetric cloning machines for qudits. In Section III
we investigate the fully
asymmetric optimal universal quantum triplicators which produce three
approximate clones with three different fidelities. Finally, Section IV 
contains a brief summary and conclusions.

\section{Asymmetric quantum duplicators}

Let us begin by briefly reviewing an isomorphism between completely positive maps
$\mathcal{S}$ and positive semidefinite operators $S \geq 0$ on the tensor
product of the input and output Hilbert spaces of map $\mathcal{S}$, denoted
respectively as $\mathcal{H}_{in}$ and $\mathcal{H}_{out}$. 
Consider a maximally entangled state on
$\mathcal{H}_{in}^{\otimes 2}$,
\begin{equation}
|\Phi^+\rangle=\frac{1}{\sqrt{d}}\sum_{j=1}^d |j\rangle|j\rangle
\end{equation}
with $d=\mathrm{dim}(\mathcal{H}_{in})$.
If the map $\mathcal{S}$ is applied to the second subsystem while nothing happens to the first one, the resulting (generally mixed) quantum state contains
all the information about the map. Qualitatively speaking, if we project
the first subsystem onto the (complex conjugate of the) input state so that
the second subsystem is projected onto the input state, then, after applying $\mathcal{S}$, it is left in the corresponding output state.
The first subsystem is therefore conventionally called the reference
system, denoted with the subscript $R$, since it keeps a memory of the state
that was processed in the channel. Mathematically, the positive semidefinite operator
\begin{equation}
S=\mathcal{I}\otimes \mathcal{S} (d\, \Phi^+_{RO})
\end{equation}
is therefore isomorphic to the map $\mathcal{S}$,
where the subscript $O$ denotes here the output system, 
$\Phi^+=|\Phi^+\rangle\langle \Phi^+|$, 
and the prefactor $d$ has been introduced for normalization purposes. 
The fact that the map $\mathcal{S}$ is trace preserving indeed implies the
condition
\begin{equation}
\mathrm{Tr}_O [S]=\openone_R.
\label{tracepreservation}
\end{equation}
The map $\mathcal{S}$ can be expressed in terms of $S$ as 
\begin{equation}
\rho \rightarrow \mathcal{S}(\rho)
= \mathrm{Tr}_{R}[\rho_R^T \otimes \openone_O \, S],
\end{equation}
where $T$ denotes the transposition in the Schmidt basis
of state $|\Phi^+\rangle$.

Let us now assume that $S$ describes the  $1\rightarrow 2$ cloning
transformation of qudits. The output Hilbert space is endowed with tensor 
product structure, $\mathcal{H}_{out}=\mathcal{H}_{A}\otimes\mathcal{H}_B$,
where the subscripts $A$ and $B$  label the two clones. For each particular
input state $|\psi\rangle$, we can calculate the fidelity of each clone as
follows,
\begin{eqnarray}
F_{A}(\psi)&=&\mathrm{Tr}(\psi_R^T\otimes \psi_{A}\otimes \openone_B \, S),
\nonumber \\
F_{B}(\psi)&=&\mathrm{Tr}(\psi_R^T\otimes \openone_{A}\otimes \psi_B \, S),
\label{FABpsi}
\end{eqnarray}
where $R$ labels the input system 
and $\psi\equiv |\psi\rangle\langle \psi|$ is a short
hand notation for the density matrix of a pure state. We are usually interested
in the average performance of the cloning machine, which can be quantified by the mean fidelities,
\begin{equation}
F_A =\int_{\psi} F_A(\psi) \, d\psi, \qquad
F_B =\int_{\psi} F_B(\psi) \, d\psi,
\label{FAB}
\end{equation}
where the measure $d\psi$ determines the kind of the cloning machines we are
dealing with. Universal cloning machines which clone equally well all states
from the input Hilbert space correspond to choosing $d\psi$ to be
the Haar measure on the group $SU(d)$. The fidelities (\ref{FAB}) 
are linear functions of the operator $S$, 
\begin{equation}
F_{A}=\mathrm{Tr}[S L_{A}], \qquad F_{B}=\mathrm{Tr}[S L_B],
\end{equation}
where the positive semidefinite operators $L_j$ are given by
\begin{equation}
L_{A}=\int_{\psi} \psi_R^T \otimes \psi_A \otimes \openone_B \, d \psi, 
\quad
L_{B}=\int_{\psi} \psi_R^T \otimes \openone_A \otimes \psi_B \, d \psi.
\end{equation}
In case of universal cloning, the integral over $d\psi$ can be easily calculated
with the help of Schur's lemma, and we get, for instance,
\begin{eqnarray*}
\int_{\psi} \psi_R^T \otimes \psi_{A} \, d \psi&=&
\frac{2}{d(d+1)}(\Pi_{RA}^{+})^{T_R} \\
&=&\frac{1}{d(d+1)}[\openone_{R}\otimes\openone_{A}+d\, \Phi_{RA}^{+}].
\end{eqnarray*}
Here, $\Pi^{+}$ denotes a projector onto symmetric subspace of two qudits, 
$d(d+1)/2$ is the dimension of this subspace, and $T_R$ stands for transposition
with respect to the subsystem $R$. Thus, we have
\begin{equation}
L_{A,B}=\frac{1}{d(d+1)}[\openone_{RAB}+d\, \tilde{L}_{A,B}] ,
\end{equation}
with
\begin{equation}
\tilde{L}_{A}=\Phi_{RA}^+\otimes \openone_{B} , 
\qquad
\tilde{L}_{B}=\Phi_{RB}^{+} \otimes \openone_A.
\end{equation}
The optimal asymmetric cloning machine $S$ should maximize a convex mixture 
of the mean fidelities $F_{A}$ and $F_{B}$ \cite{Fiurasek03,Lamoureux04},
\begin{equation}
F=pF_{A}+(1-p)F_{B}=\mathrm{Tr}[SL],
\label{convexmixture}
\end{equation}
where $L=pL_A+(1-p)L_{B}$ and $p$ is a parameter that controls the asymmetry of
the cloner.  The maximization of $F$ for a given value of $p$ can be
equivalently rephrased as a maximization of $F_{B}$ for a fixed value of $F_A$.
Suppose that we find $S$ that maximizes $F$. It is then clear that for a given
$F_{A}$ this map yields maximum possible $F_B$, because any higher $F_B$ would
increase $F$. This explains why optimal asymmetric cloners can be found simply 
by maximizing the convex mixture of single-clone fidelities with variable mixing
ratio.

The maximum achievable $F$ is upper bounded by the maximum 
eigenvalue $\lambda_{max}$ of the operator $L$ \cite{Fiurasek01}. Taking into account the trace-preservation condition, we have 
\begin{equation}
F \leq d \, \lambda_{max}.
\end{equation}
Although this bound need not be saturated in general \cite{Fiurasek01,Fiurasek02},
it is reached by the optimal asymmetric $1\rightarrow 2$ universal cloning 
machines, as we shall show below. 
It follows that we have to calculate the eigenvalues of the operator
\begin{equation}
L=\frac{1}{d(d+1)}[\openone_{RAB}+ d\, \tilde{L}]
\end{equation}
with 
\begin{equation}
\tilde{L}=p\, \tilde{L}_A+ (1-p)\, \tilde{L}_B .
\end{equation}
We can neglect the trivial part of $L$ which is proportional to the identity
operator, and only need to investigate the eigenstates and eigenvalues
of $\tilde{L}$.
Luckily, this problem is greatly simplified by noting that $\tilde{L}$
has a support of dimension $2d$, spanned by $|\Phi^+\rangle_{RA}|k\rangle_B$ and
$|\Phi^+\rangle_{RB}|k\rangle_A$. This implies that $\tilde{L}$ has at most $2d$
non-zero eigenvalues. Moreover, it turns out that there are only two 
$d$-fold degenerate eigenvalues, $\lambda_1$ and $\lambda_2$.
The eigenstates have the following form,
\begin{equation}
|\lambda_j;k\rangle=\alpha\, |\Phi^+\rangle_{RA}|k\rangle_{B}
+\beta\, |\Phi^+\rangle_{RB}|k\rangle_{A},
\label{rtilde}
\end{equation}
where $j=1,2$ and $k=1,\cdots , d$.
The two eigenvalues $\lambda_{1}> \lambda_2$ are roots of the quadratic equation
\begin{equation}
\lambda^2-\lambda+p(1-p)[1-d^{-2}]=0
\end{equation}
and the ratio $\beta/\alpha$, which fixes the eigenstate (\ref{rtilde}), 
can be expressed in terms of $\lambda$, $p$, and $d$ as
\begin{equation}
\frac{\beta}{\alpha}=d(\lambda/p-1).
\label{abratio}
\end{equation}
Since $\lambda$ is real, we can assume without loss of generality that 
$\alpha$ and $\beta$ are both real and $\alpha \geq 0$.
By properly normalizing the eigenstates $|\lambda_j;k\rangle$, we get
\begin{equation}
\alpha^2+\beta^2+\frac{2\alpha\beta}{d}=1.
\label{alphabetanorm}
\end{equation}

The optimal cloning transformation $S$ is then simply the projector onto the
$d$-dimensional sub-space spanned by the eigenstates $|\lambda_1;k\rangle$
corresponding to the maximum eigenvalue $\lambda_1$,
\begin{equation}
S=\sum_{k=1}^d|\lambda_1;k\rangle\langle \lambda_1;k|.
\label{Sduplicatoroptimal}
\end{equation}
Note that $\lambda_1>p$ hence both $\alpha$ and $\beta$ in Eq. (\ref{rtilde})
are positive.
One can easily check that $\mathrm{Tr}_{AB}[S]=\openone_{R}$, hence $S$ is a
trace-preserving map.

Moreover, $F=d \, \lambda_{max}$ by construction, which proves the 
optimality. The fidelities of the optimal clones $A$ and $B$ can be 
obtained in terms of the coefficients $\alpha$ and $\beta$ by noting first that
\begin{eqnarray}
\langle \lambda_1;k| \tilde{L}_A \,|\lambda_1;k\rangle = (\alpha+\beta/d)^2 , \nonumber \\
\langle \lambda_1;k| \tilde{L}_B \,|\lambda_1;k\rangle = (\beta+\alpha/d)^2 ,
\end{eqnarray}
so that, using Eq.~(\ref{alphabetanorm}), we get
\begin{equation}
\mathrm{Tr}[S \tilde{L}_{A}] = d-\frac{d^2-1}{d}\, \beta^2, \quad
\mathrm{Tr}[S \tilde{L}_{B}] = d-\frac{d^2-1}{d}\, \alpha^2.
\end{equation}
Therefore, we obtain for the fidelities of the asymmetric cloner
\begin{equation}
F_A=1-\frac{d-1}{d}\beta^2, \qquad F_B=1-\frac{d-1}{d}\alpha^2,
\label{fidelitiesAB}
\end{equation}
where $\alpha^2$ and $\beta^2$ are the so-called 
depolarizing fractions as discussed in Ref. \cite{Cerf98}. 
The expressions (\ref{alphabetanorm}) and (\ref{fidelitiesAB}) 
exactly coincide with the formula
characterizing the class of asymmetric cloning machines 
derived in \cite{Cerf98}, which therefore is optimal.
 
The optimal cloning  map (\ref{Sduplicatoroptimal}) can be realized 
unitarily by purifying $S$ into the state
\begin{equation}
|\Phi\rangle =
\alpha \, |\Phi^+\rangle_{RA} |\Phi^+\rangle_{BE}
+ \beta \, |\Phi^+\rangle_{RB} |\Phi^+\rangle_{AE},
\end{equation}
where $E$ stands for an ancillary system, that is, we get $S$
when tracing over $E$. The resulting isometry that transforms the input
single-qudit state $|\psi\rangle$ onto the output state of three qudits 
(two clones and one anti-clone) can be written, by projecting the 
reference system $R$ onto $|\psi^*\rangle$, as
\begin{equation}
|\psi\rangle \rightarrow 
\alpha \, |\psi\rangle_{A}|\Phi^+\rangle_{BE}
+\beta \, |\psi\rangle_{B}|\Phi^+\rangle_{AE}.
\end{equation}

\section{Asymmetric quantum triplicators}

Having proved the optimality of the universal asymmetric $1\rightarrow 2$
cloning machines, we now use the same techniques to construct the optimal
universal asymmetric $1\rightarrow 3$ cloners. These machines produce three
clones, $A$, $B$, and $C$, each clone possibly having a different fidelity ($F_A$, $F_B$, and $F_C$). 
The optimal asymmetric cloning machine  should  maximize the 
cloning fidelities such that for a given pair of fidelities (say $F_A$ and
$F_B$) the fidelity of the third clone ($F_C$) is maximum. 

The output Hilbert space of the asymmetric quantum triplicator is a tensor
product of Hilbert spaces of the three clones. The average  fidelity of $j$th
clone can be again expressed as $F_j=\mathrm{Tr}[S L_j]$ with $j \in \{A,B,C\}$,
 where now
\begin{equation}
L_{A}=\frac{1}{d(d+1)}[\openone_{R}\otimes\openone_{A}+d
\Phi_{RA}^{+}]\otimes\openone_{BC},
\end{equation}
where $R$ indicates the reference, and $L_{B}$ and $L_{C}$ can be obtained by cyclic permutation of $A,B,C$. In analogy with Eq.~(\ref{convexmixture}),
the optimal asymmetric $1\rightarrow 3$ cloning machine should maximize a
convex combination of the three single-clone fidelities,
\begin{equation}
F=a F_{A}+b F_{B} + cF_{C},
\label{Fabc}
\end{equation}
where $a+b+c=1$, $a,b,c\geq 0$ and the asymmetry of the cloner is determined by
the ratios $a/b$ and $a/c$. The fidelity (\ref{Fabc}) can be rewritten as 
$F=\mathrm{Tr}[S L]$, where $L=a L_A+b L_{B}+c L_C$. Similarly 
as in the case of $1\rightarrow 2$ cloning, we have to determine the eigenspace
corresponding to the maximum eigenvalue of 
\begin{eqnarray}
L&=&\frac{1}{d(d+1)} \left[\openone_{RABC}
+d \, \tilde{L} \right],
\label{Rabc}
\end{eqnarray}
where 
\begin{equation}
\tilde{L}=
a\, \Phi_{RA}^{+}\otimes\openone_{BC}+
b\, \Phi_{RB}^{+}\otimes\openone_{AC}+
c\, \Phi_{RC}^{+}\otimes\openone_{AB}.
\end{equation}
Due to the high symmetry, the operator $\tilde{L}$ has only 
six different non-zero eigenvalues. Three of them are $d(d+1)/2$-fold 
degenerate and the corresponding eigenstates read,
\begin{eqnarray}
|\lambda_+;kl\rangle&=&
\alpha\, |\Phi^+\rangle_{RA}|kl^{+}\rangle_{BC} 
 +\beta\, |\Phi^+\rangle_{RB}|kl^{+}\rangle_{AC} \nonumber \\
&&+\gamma\, |\Phi^+\rangle_{RC}|kl^{+}\rangle_{AB},
\label{rklplus}
\end{eqnarray}
with $l \geq k$. Here we take $|kl^{+}\rangle=(|kl\rangle+|lk\rangle)/\sqrt{2}$
if $k\ne l$, while $|kk^{+}\rangle=|kk\rangle$. The three 
eigenvalues ($\lambda_{+,1}>\lambda_{+,2}>\lambda_{+,3}$) can be determined 
as roots of the cubic equation
\begin{eqnarray}
P_{+}(\lambda_+) &\equiv& \lambda_{+}^3-\lambda_+^2
+\lambda_+(ab+bc+ac)\left(1-d^{-2}\right) \nonumber \\
&&-abc\left(1+2d^{-3}-3d^{-2}\right)=0,
\end{eqnarray}
and the coefficients $\alpha,\beta,\gamma$ can be expressed in terms of $a,b,c$,
and $\lambda_{+}$ by solving the system of linear equations
\begin{eqnarray}
(\lambda_{+}-a)\alpha-\frac{a}{d}(\beta+\gamma)=0, \nonumber \\
(\lambda_{+}-b)\beta-\frac{b}{d}(\alpha+\gamma)=0, \nonumber \\
(\lambda_{+}-c)\gamma-\frac{c}{d}(\alpha+\beta)=0 .
\label{alfabetagamma}
\end{eqnarray} 
The normalization of the eigenstate $|\lambda_+;kl\rangle$ 
imposes the constraint
\begin{equation}
\alpha^2+\beta^2+\gamma^2+\frac{2}{d}(\alpha\beta+\alpha\gamma+\beta\gamma)=1.
\label{alfabetagammanorm}
\end{equation}
The other three eigenvalues correspond to the anti-symmetric combinations of
$|kl\rangle$ and $|lk\rangle$, that is, $|kl^{-}\rangle=(|kl\rangle-|lk\rangle)/\sqrt{2}$,
and are thus $d(d-1)/2$-fold degenerate. 
The eigenstates are given by
\begin{eqnarray}
|\lambda_-;kl\rangle&=&
\alpha|\Phi^+\rangle_{RA}|kl^{-}\rangle_{BC}
+\beta|\Phi^+\rangle_{RB}|kl^{-}\rangle_{AC}  \nonumber \\
& & +\gamma|\Phi^+\rangle_{RC}|kl^{-}\rangle_{AB},
\end{eqnarray}
with $l >k$, and the cubic equation for the eigenvalues $\lambda_-$ reads
\begin{eqnarray}
P_{-}(\lambda_-)&\equiv& \lambda_{-}^3-\lambda_{-}^2
+\lambda_{-}(ab+bc+ac)\left(1-d^{-2}\right) \nonumber \\
& &-abc\left(1-2d^{-3}-3d^{-2}\right)=0.
\end{eqnarray}
Since the polynomials $P_{+}(\lambda)$ and $P_{-}(\lambda)$ differ only
in their zeroth order terms, their graphs look identical up to a vertical
shift of $4abc/d^3$.
This simple geometrical observation reveals that the maximum eigenvalue
$\lambda_{+,1}$ is always larger than the maximum eigenvalue
$\lambda_{-,1}$. Hence, in determining the optimal cloning transformation, 
which corresponds to the maximum eigenvalue of (\ref{Rabc}),
we have to consider only 
the eigenstates (\ref{rklplus}). It follows from the structure of the operator $\tilde{L}$
that $\lambda_{+,1}\geq \mathrm{max}(a,b,c)$. This, together with 
Eq. (\ref{alfabetagamma}) implies that the coefficients $\alpha$, $\beta$,
and $\gamma$ of an eigenstate corresponding to 
the maximum eigenvalue $\lambda_{+, 1}$
must be all positive (or all negative). The optimal trace-preserving 
$1\to3$ cloning map can then be expressed simply 
as the properly normalized projector onto the subspace 
spanned by the $d(d+1)/2$ eigenstates (\ref{rklplus}) with eigenvalue $\lambda_{+,1}$,
\begin{equation}
S= \frac{2}{d+1}\sum_{l \geq k} 
|\lambda_{+,1};kl\rangle\langle \lambda_{+,1};kl|,
\end{equation}
where the prefactor originates from the constraint that $\mathrm{Tr}(S)=d$.
A unitary implementation of this CP map requires two ancilla systems, $E$ and
$F$, and can be characterized by the purification of $S$, namely
\begin{eqnarray}
|\Phi\rangle &=&
\sqrt{d} \,\mathcal{C}\left[ 
\alpha\, |\Phi^+\rangle_{RA}(|\Phi^{+}\rangle_{BE}|\Phi^{+}\rangle_{CF}
+|\Phi^{+}\rangle_{BF}|\Phi^{+}\rangle_{CE}) \right. \nonumber \\
&&+\beta\, |\Phi^+\rangle_{RB}(|\Phi^{+}\rangle_{AE}|\Phi^{+}\rangle_{CF}
+|\Phi^{+}\rangle_{AF}|\Phi^{+}\rangle_{CE}) \nonumber \\
&&\left.+\gamma\, |\Phi^+\rangle_{RC}(|\Phi^{+}\rangle_{AE}|\Phi^{+}\rangle_{BF}
+|\Phi^{+}\rangle_{AF}|\Phi^{+}\rangle_{BE}) \right], \nonumber 
\end{eqnarray}
where we have used the identity
\begin{eqnarray}
\lefteqn{ \frac{2}{d}\sum_{l\ge k} |kl^+\rangle_{BC} |kl^+\rangle_{EF} = }
\hspace{1cm} \nonumber \\
&& |\Phi^+\rangle_{BE} |\Phi^+\rangle_{CF} 
+ |\Phi^+\rangle_{BF} |\Phi^+\rangle_{CE},
\end{eqnarray}
and the normalization constant is $\mathcal{C}=\sqrt{d/(2(d+1))}$.
Therefore, by projecting $R$ onto $|\psi^*\rangle$, we see that
any pure input state $|\psi\rangle$ transforms according to
\begin{eqnarray}
|\psi\rangle &\rightarrow &
\mathcal{C}\left[ \alpha
|\psi\rangle_{A}(|\Phi^{+}\rangle_{BE}|\Phi^{+}\rangle_{CF}
+|\Phi^{+}\rangle_{BF}|\Phi^{+}\rangle_{CE}) \right. \nonumber \\
&&+\beta|\psi\rangle_{B}(|\Phi^{+}\rangle_{AE}|\Phi^{+}\rangle_{CF}
+|\Phi^{+}\rangle_{AF}|\Phi^{+}\rangle_{CE}) \nonumber \\
&&\left.+\gamma |\psi\rangle_{C}(|\Phi^{+}\rangle_{AE}|\Phi^{+}\rangle_{BF}
+|\Phi^{+}\rangle_{AF}|\Phi^{+}\rangle_{BE}) \right]. \nonumber 
\end{eqnarray}
It can be easily verified that this transformation is universal, i.e. the
single-clone fidelities do not depend on the output state.

\begin{figure}[!t!]
\centerline{\psfig{figure=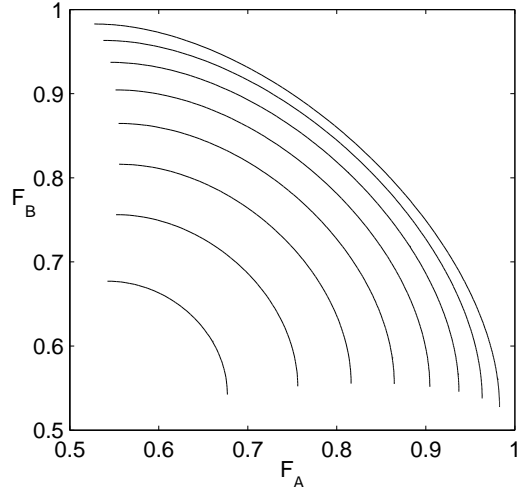,width=0.8\linewidth}}
\caption{The trade-off between the fidelities $F_A$ and $F_B$ for a fixed
fidelity $F_C$ is shown for the optimal
universal asymmetric $1\rightarrow 3$ cloning of qubits.
The curves are plotted for several different values of 
$F_C(n)=0.6+0.05 n$, $n=0,\ldots,7$, the most inward curve corresponding 
to the highest value of $F_C$.}
\end{figure}

We can express the
fidelities in terms of the coefficients $\alpha$, $\beta$, and $\gamma$,
by noting that
\begin{eqnarray}
\langle \lambda_{+,1};kl| \Phi^+_{RA} \otimes \openone_{BC}
|\lambda_{+,1};kl\rangle = (\alpha+\beta/d+\gamma/d)^2 , \nonumber \\
\langle \lambda_{+,1};kl| \Phi^+_{RB} \otimes \openone_{AC}
|\lambda_{+,1};kl\rangle = (\beta+\alpha/d+\gamma/d)^2 , \nonumber \\
\langle \lambda_{+,1};kl| \Phi^+_{RC} \otimes \openone_{AB}
|\lambda_{+,1};kl\rangle = (\gamma+\alpha/d+\beta/d)^2 .
\end{eqnarray}
Using the normalization condition (\ref{alfabetagammanorm}), we obtain
the fidelity triplet
\begin{eqnarray}
F_{A}=1-\frac{d-1}{d}\left[\beta^2+\gamma^2+\frac{2\beta\gamma}{d+1}\right], \nonumber \\
F_{B}=1-\frac{d-1}{d}\left[\alpha^2+\gamma^2+\frac{2\alpha\gamma}{d+1}\right], \nonumber \\
F_{C}=1-\frac{d-1}{d}\left[\alpha^2+\beta^2+\frac{2\alpha\beta}{d+1}\right].
\end{eqnarray}
This, together with the normalization condition (\ref{alfabetagammanorm}) and the constraints $\alpha
\geq0, \beta \geq0,\gamma\geq 0 $,
provides a parametric description of the whole
class of the optimal universal asymmetric $1\rightarrow 3$ cloning machines
in a Hilbert space of arbitrary dimension $d$.

As an example, in Fig. 1 we plot the trade-off between $F_A$ and $F_B$ for
several different values of the fidelity of the third clone $F_C$
for $1\rightarrow 3$ asymmetric cloning of qubits, $d=2$. 
Note, that in the limit where one of the three coefficients 
$\alpha,\beta,\gamma$ is equal to zero the asymmetric $1\rightarrow 3$ 
cloning essentially reduces to the optimal asymmetric $1\rightarrow 2$ cloning. 
However, even in this case the fidelity of
the third clone is larger than $1/2$, which is what one could have 
naively expected.
This interesting effect is clearly visible in Fig. 1. The endpoints of the 
curves showing the trade-off between $F_A$ and $F_B$ for a fixed $F_C$
correspond to optimal $1\rightarrow 2$ asymmetric cloning in the subspace of
qubits $A$ and $C$  (or $B$ and $C$). Note that the endpoints do not lie on the 
line $F_B=1/2$ ($F_A=1/2$) and the fidelity $F_B$ ($F_A$) is thus 
higher than $1/2$ even in this limit case. This behavior can be easily understood by noting that in
the $1\rightarrow 2$ cloning, the ancilla (anti-clone) carries some information
about the input and a third clone with fidelity larger than $1/2$ 
can be produced simply by applying the optimal approximate universal-NOT gate 
\cite{Gisin99,Buzek99unot} to the anti-clone. In particular, for 
$\alpha=\beta=1/\sqrt{3}$ and $\gamma=0$ we obtain the optimal triplet of
fidelities $F_A=F_B=5/6$ and $F_C=5/9$. The three clones exhibiting these
fidelities can be prepared by first performing the optimal
symmetric $1\rightarrow 2$ universal cloning which produces two clones with
fidelity $5/6$. The third clone is then obtained from the anti-clone 
by applying the approximate UNOT which yields a clone with fidelity exactly
$5/9$.

\section{Conclusions}

In summary, we have investigated  asymmetric universal cloning in arbitrary dimension.
We have proved the optimality of the universal asymmetric $1\rightarrow 2$ 
cloning machines that have been previously considered as  possible efficient
attacks on certain classes of quantum key distribution protocols. We have then
extended the concept of asymmetric cloning to quantum triplicators, which
produce three clones of different fidelity. We have
derived a simple parametric description of the  optimal asymmetric $1\rightarrow
3$ cloning machines and we have provided an explicit formula for the optimal 
cloning transformation.

We anticipate that
our results may play an important role in quantum information theory, for
instance in the analysis of quantum information distribution in quantum networks 
and in studies of eavesdropping strategies on multi-party quantum communication 
protocols. 

\begin{acknowledgments}

We acknowledge financial support from the EU under project SECOQC
(IST-2002-506813) . JF and RF also acknowledge support
from the  grant MSM 6198959213 of the Czech Ministry of Education.
NJC acknowledges financial support from the Communaut\'e Fran\c{c}aise de
Belgique under grant ARC 00/05-251 and from the IUAP programme of the Belgian
government under grant V-18.

\end{acknowledgments}

\end{document}